\documentclass[12pt]{article}
\pdfoutput=1

\usepackage{hyperref}
\hypersetup{
  colorlinks=true,
  pdfborder={0 0 0},
  linktocpage=false
}

\usepackage[normalem]{ulem}
\usepackage[utf8]{inputenc}
\usepackage[left=2.55cm, right=2.55cm, top=2.55cm, bottom=2.55cm]{geometry}
\usepackage{amsmath,amssymb}
\usepackage{slashed}
\usepackage{xcolor}
\usepackage{graphicx}
\usepackage{url}
\usepackage{cancel}
\usepackage{cite}
\usepackage{tabularx,booktabs}
\usepackage{multicol}
\usepackage{units}
\usepackage{xspace}
\usepackage[labelfont=bf]{caption}
\usepackage{mathrsfs}

\definecolor{darkred}{rgb}{0.6,0,0}
\definecolor{darkpurple}{rgb}{0.5,0,0.5}

\def\gsim{\raise0.3ex\hbox{$\;>$\kern-0.75em\raise-1.1ex\hbox{$\sim\;$}}}
\def\lsim{\raise0.3ex\hbox{$\;<$\kern-0.75em\raise-1.1ex\hbox{$\sim\;$}}}

\begin{document}

\begin{center}
\vspace*{15mm}

\vspace{1cm}
{\Large \bf 
Effective Field Theory and Scalar Triplet Dark Matter} \\
\vspace{1cm}

{\bf Carolina Arbel\'aez$^{\text{a}}$, Marcela Gonz\'alez$^{\text{b}}$,
  Martin Hirsch$^{\text{c}}$, Nicol\'as A. Neill$^{\text{d}}$,\\ 
  Diego Restrepo$^{\text{e}}$ }

\vspace*{.5cm} $^{\text{a}}$  Universidad T\'ecnica Federico Santa Mar\'ia
 and Centro Cient\'ifico Tecnol\'ogico\\
de Valpara\'iso CCTVal, Casilla 110-V, Valpara\'iso, Chile

\vspace*{.5cm} $^{\text{b}}$  Instituto de F\'isica y Astronom\'ia,\\
Universidad de Valpara\'iso,\\ Avenida Gran Breta\~na 1111, Valpara\'iso, Chile

\vspace*{.5cm} $^{\text{c}}$Instituto de F\'{\i}sica Corpuscular
(CSIC-Universitat de Val\`{e}ncia), \\ C/ Catedr\'atico Jos\'e
Beltr\'an 2, E-46980 Paterna (Val\`{e}ncia), Spain

\vspace*{.5cm} $^{\text{d}}$ Departamento de Ingenier\'ia El\'ectrica-Electr\'onica, \\ Universidad de Tarapac\'a, Arica, Chile

\vspace*{.5cm} $^{\text{e}}$ Instituto de F\'isica, \\ Universidad de Antioquia, Calle 70 No 52-21, Medell\'in, Colombia

 \vspace*{.3cm}
 \href{mailto:carolina.arbelaez@usm.cl}{carolina.arbelaez@usm.cl},
 \href{mailto:marcela.gonzalezpi@uv.cl}{marcela.gonzalezpi@uv.cl},
\href{mailto:mahirsch@ific.uv.es}{mahirsch@ific.uv.es},
 \href{mailto:naneill@outlook.com}{naneill@outlook.com},
 \href{mailto:restrepo@udea.edu.co}{restrepo@udea.edu.co}
\end{center}

\vspace*{10mm}
\begin{abstract}

We discuss an extension of the standard model with a real scalar
triplet, $T$, including non-renormalizable operators (NROs) up to
$d=6$. If $T$ is odd under a $Z_2$ symmetry, the neutral component of $T$
is a good candidate for the dark matter (DM) of the universe. We
calculate the relic density and constraints from direct and indirect
detection on such a setup, concentrating on the differences with
respect to the simple model for a DM $T$ with only renormalizable
interactions. Bosonic operators can change the relic density of
the triplet drastically, opening up new parameter space for the
model. Indirect detection constraints, on the other hand, rule
out an interesting part of the allowed parameter space already today
and future CTA data will, very likely, provide a decisive test 
for this setup. 

\end{abstract}

\section{Introduction}\label{sec:introduction}

Among the long list of proposed explanations for the dark matter (DM)
in the universe, weakly interacting massive particles (WIMPs), are
probably the solution which has been studied most. For reviews on
WIMPs see, for example
\cite{Bertone:2004pz,Arcadi:2017kky,Roszkowski:2017nbc}. WIMPs are
attractive candidates for a variety of reasons, not least among 
them the fact that their thermal production in the early universe 
makes it possible -- in principle -- to calculate definite predictions 
for cross sections for experiments searching for signals of DM 
today. 

Any DM candidate must be able to reproduce the relic density (RD) as
inferred from PLANCK data \cite{Planck:2018vyg} and also obey a number
of search constraints. For WIMPs, the most important bounds are those
from two types of searches. First, there are the direct detection (DD)
experiments, like LUX-ZEPLIN (LZ) \cite{LZ:2022lsv,LZCollaboration:2024lux},
and XENON-nt \cite{XENON:2023cxc}. Future DD experiments, such as
DARWIN \cite{DARWIN:2016hyl}, XLZD \cite{Aalbers:2022dzr,XLZD:2024nsu}
or also PandX-xt \cite{PandaX:2024oxq} will improve existing limits by
(1-2) orders of magnitude. These constraints limit the list of
possible WIMP candidates already today to a finite number of
candidates \cite{Bottaro:2021snn,Bottaro:2022one}, and may exclude
many more DM candidates (or discover the correct one!) in the future.
Second, also indirect detection (ID) constraints\footnote{For 
a recent discussion see for example~\cite{Cirelli:2024ssz,Beauchesne:2024tuc}.} 
can be important. Existing data from H.E.S.S collaboration for 546~h
\cite{HESS:2022ygk,Montanari:2023bzn} already rule out interesting
portions of parameter space, while the future CTA data
\cite{CTA:2020qlo,CTAO:2024wvb} is expected to lead to much stronger
constraints.

The neutral component of a scalar triplet with hypercharge zero, $T$,
can be a good WIMP dark matter candidate
\cite{Cirelli:2005uq,Cirelli:2007xd}.  A real $T$ couples at the
renormalizable level only to the standard model (SM) Higgs and this simple setup is
quite predictive, since the relic DM density and the direct detection
cross section essentially depend only on two parameters: The mass of
$T^0$ and the quartic coupling of $T$ to the SM Higgs. This minimal,
renormalizable setup is, however, nowadays on the verge of being
excluded by H.E.S.S data \cite{HESS:2022ygk,Montanari:2023bzn} and
will definitely be either discovered (or excluded) in the
not-too-distant future, see the discussion in section \ref{sec:NS}.

\begin{figure}[t]
    \centering
    \includegraphics[scale=0.6]{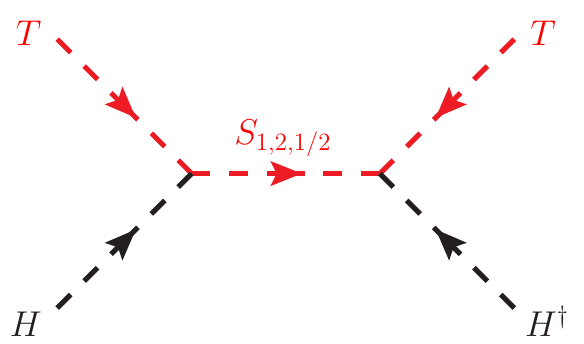}\hskip7mm
    \includegraphics[scale=0.6]{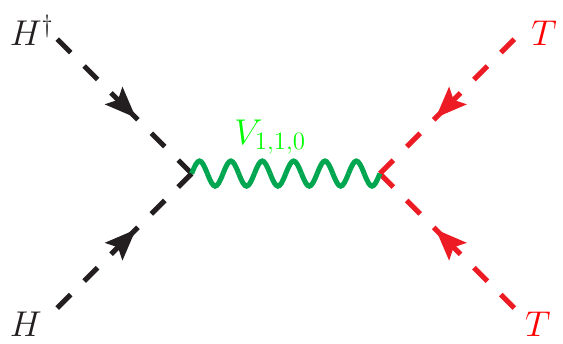}
    \caption{Co-annihilation diagram (left) for a model with dark
      matter triplet and an additional heavy (inert) scalar,
      $S_{1,2,1/2}$. This scalar is $Z_2$ odd. s-channel annihilation
      of the dark matter triplet (right) in a model with an additional
      vector, $Z'=V_{1,1,0}$. This vector is $Z_2$ even. In this setup
      the $Z_2$ could be a remnant of the symmetry breaking that gave
      the $Z'$ its mass.}
    \label{fig:UVExas}
\end{figure}

However, in many models beyond the SM (BSM), the dark matter candidate
is just the lightest particle of a possibly much more complicated
``dark sector''.\footnote{The best known example for such a case is
  supersymmetry: In the MSSM (``minimal supersymmetric standard
  model'') all particles of the SM have a superpartner, odd under
  R-parity, the DM is just the lightest particle in the SUSY
  spectrum.} These heavier dark sector particles can have important
effects on the DM properties. For example, the relic density can
change due to co-annihilation diagrams or $s$-channel resonances.  One
example diagram for each is shown for a toy model in
fig. (\ref{fig:UVExas}).  Here, $T$ is assumed to be to lightest $Z_2$
odd particle. On the left, we assume there is a second, heavier Higgs
doublet, $S_{1,2,1/2}$, also odd under $Z_2$.\footnote{If
  $S_{1,2,1/2}$ is the lightest odd particle, it can itself be the
  dark matter, the well-known inert doublet model (IDM)
  \cite{LopezHonorez:2006gr}.} On the right, we added a
$Z'=V_{1,1,0}$. This could be the vector of an extended gauge group,
broken in the ultra-violet in such a way that a global $Z_2$
remains. Such an ansatz could explain the origin of the $Z_2$,
otherwise introduced by hand.

On the other hand, both toy models generate the same effective
dimension six ($d=6$) operator, ${\cal O}_{3} \propto
(H^{\dagger}D_{\mu}H)(TD^{\mu}T)$ in the limit where the scalar or
vector mass are larger than $m_T$. In the case there is some hierarchy
in the masses of the dark sector, effective field theories are
therefore a useful tool for model-independent studies of dark matter
phenomenology.

Indeed, application of EFTs for dark matter calculations has already
gained some momentum in the past few years. For example,
\cite{Bishara:2017pfq,Brod:2017bsw} studied WIMP direct detection
using low-energy EFT.  Recently, \cite{Criado:2021trs} has given the
complete set of operators up to dimension six ($d=6$) for the case of
models in which one dark matter particle (either scalar, fermion or
vector) is added to the SM particle content. The paper
\cite{Criado:2021trs} then studies in detail the dark matter
phenomenology for a singlet scalar DM, $S_{1,1,0}$, and for heavy
neutral lepton DM, $L'=F_{1,2,1/2}$.  \cite{Coito:2022kif} used a
sterile neutrino portal for a fermionic singlet DM, including
non-renormalizable operators (NROs).
Ref \cite{Aebischer:2022wnl} adds singlets (either scalars, fermions 
or vectors) with NRO up to $d=6$, and discusses how models with 
two dark vectors can explain dark matter via freeze-in. And there 
is the work of \cite{Liang:2023yta} which adds up to two DM particles 
and writes down the low-energy EFT operators up to $d=7$. The authors 
call this ``DSEFT'', dark sector effective field theory.

In our current work, we will study the phenomenology of a dark matter
$T$ adding NROs. We will call this framework TSMEFT, i.e. the
``standard model effective field theory supplemented by a scalar
triplet''. In the next section, we will briefly introduce the model
lagrangian. Section \ref{relic} discusses, how the different, possible
$d=6$ operators affect the relic density of the dark matter
candidate. In section \ref{sec:cnst} we discuss, how operators are
constrained from direct and indirect detection experiments and discuss
briefly, which data will be important for us. Section \ref{sec:NS}
then shows the main resuls of this paper. There, we will discuss our
numerical results and show, which operators are most (and least)
constrained at present for a DM model with a hyperchargeless triplet.
We then close with a short summary and conclusion. For comepleteness,
in appendix \ref{subsec:OddOps} we list for also operators odd in $T$,
while in \ref{subsec:CmplOps} we discuss possible, additional
operators for a complex $T$. Finally, in appendix \ref{subsec:UV} we
list possible tree-level UV completions for the operators studied in
this paper.

\section{Scalar triplet lagrangian up to $d=6$} \label{sec:tsmeft}

We add a scalar $S_{1,3,0}=T$ to the SM particle content. We will
consider the simpler case of $T$ being real. The renormalizable
Lagrangian of the model consists of the SM Lagrangian, plus the
following new terms:
\begin{equation}\label{eq:Lag}
{\cal L}^{\rm TSM} = {\cal L}^{\rm SM} + \frac{1}{2}m_T^2 T^2  
+ \mu_T (H^{\dagger}H) T + {\rm h.c.} +  
\frac{1}{2}\lambda_{HT} H^{\dagger}H T^2 - \frac{1}{4}\lambda_T T^4.
\end{equation}
The term proportional to $\mu_T$ allows the field $T$ to decay to SM
particles. We will forbid this term and all other terms odd in $T$ 
by assuming a $Z_2$ symmetry under which the only odd particle is 
$T$. 

Apart from the renormalizable lagrangian, eq. (\ref{eq:Lag}), we
will add non-renormalizable operators up to dimension 6:
\begin{equation}
  {\cal L} = {\cal L}^{\rm TSM}
    + \sum_{i} \frac{C_{i}}{\Lambda^{d-4}} {\cal O}_{i} .
\end{equation}
Table \ref{tab:ops6} lists all possible effective operators (up to
$d=6$), which contain an even number of $T$'s. We do not write the 
Wilson coefficients explicitly, for operators with standard model 
fermions the coefficients are matrices in flavour space. The 
table defines the operators, gives the number of free parameters 
in the Wilson coefficients and whether the operator is self-conjugate 
or not. For completenes, in appendix \ref{subsec:OddOps} we also 
give operators odd in $T$ up to $d=6$.

\begin{table}[h]
\centering
\begin{tabular}{cccc}
\hline
Name & Operator & $\#$ parameters & +h.c.? \\
\hline
\hline
${\cal O}_{1}$ & $(\overline{L}\gamma_{\mu}L)(TD^{\mu}T)$ & 9   & no   \\
${\cal O}_{2}$ & $(\overline{Q}\gamma_{\mu}Q)(TD^{\mu}T)$ & 9   & no   \\
${\cal O}_{3}$ & $(H^{\dagger}D_{\mu}H)(TD^{\mu}T)$  & 2   & no   \\
${\cal O}_{4}$ & $\overline{L}e_RHTT$ & 18   & yes   \\
${\cal O}_{5}$ & $\overline{Q}d_RHTT$ & 18   & yes   \\
${\cal O}_{6}$ & $\overline{Q}u_RH^{\dagger}TT$ & 18   & yes   \\
${\cal O}_{7}$ & $B^{\mu\nu}B_{\mu\nu}TT$ & 1   & no   \\
${\cal O}_{8}$ & $W^{\mu\nu}W_{\mu\nu}TT$ & 1   & no   \\
${\cal O}_{9}$ & $G^{\mu\nu}G_{\mu\nu}TT$ & 1   & no   \\
${\cal O}_{10}$ & $H^{\dagger}H^{\dagger}HHTT$ & 2   & no   \\
${\cal O}_{11}$ & $H^{\dagger}HTTTT$ & 1   & no   \\
${\cal O}_{12}$ & $(TD^{\mu}T)(TD_{\mu}T)$ & 1   & no   \\
${\cal O}_{13}$ & $T^6$ & 1   & no   \\
\hline
\end{tabular}
\caption{Operators even in $T$ in TSMEFT at $d=6$.  This set of operators 
is allowed for a $T$ odd under $Z_2$. Operators ${\cal O}_{10}-{\cal O}_{13}$ 
are listed only for completeness, they are less important for the 
phenomenology studied in this paper.}
\label{tab:ops6}
\end{table}

Not all of the operators in table \ref{tab:ops6} will be important for 
dark matter phenomenology. Operators ${\cal O}_{11}-{\cal O}_{13}$ 
do not lead to any change in the number density of $T$ in the early 
universe and we have therefore not included them in our numerical 
studies. Also, ${\cal O}_{10}$ is a six-particle operator, sub-dominant 
to ${\cal O}_{3}$ in its effects on DM phenomenology. We do not study 
this operator in detail either.

Numerically, we only calculate the effects of NROs. However, one should keep in
mind that there are special kinematic configurations, for
which the use of EFTs is {\em not} well justified. The best example
for this is probably the s-channel diagram on the right of fig.
(\ref{fig:UVExas}) in the special case where $m_T \simeq m_{Z'}/2$. In
this case the cross section will be resonantly enhanced and this
effect is not captured by the EFT. It may therefore be of interest in
some cases to go to model-dependent studies. While this is beyond the
scope of the present work, in appendix \ref{subsec:UV}, we give all
possible tree-level completions for operators ${\cal O}_{1} - 
{\cal O}_{9}$ for completeness.  Recall, two example diagrams for toy
UV models for ${\cal O}_{3}$ were already discussed briefly in the
introduction.

\section{Relic density}\label{relic}

In this section we study the effects of the operators ${\cal O}_{1} -
{\cal O}_{9}$ on the DM relic abundance, $\Omega h^2$. The operators
are listed in table \ref{tab:ops6}.  We have implemented all nine
operators in \texttt{CalcHEP} format \cite{Pukhov:2004ca} using
\texttt{FeynRules} \cite{Alloul:2013bka} and calculated the relic
density of DM numerically using \texttt{MicrOMEGAs}
\cite{Belanger:2013oya,Belanger:2018ccd,Belanger:2020gnr}.
Results for all operators will be discussed in section \ref{sec:NS}. 

However, let us first discuss some basics of DM freeze-out which will
allow us to understand the numerical results qualitatively. The DM
relic abundance $\Omega h^2$ generated within the standard thermal
WIMP paradigm depends of the thermal average of the DM annihilation
cross section times the relative DM velocity, $\left < \sigma v
\right>$. It can be written as \cite{Griest:1990kh}:

\begin{equation}
  \Omega h^2 = \frac{1.09 \times 10^9\,\text{GeV}^{-1}}{g_{*}^{1/2}m_{\text{Pl}}}
  \frac{1}{J(x_{f})},
  \;
  J(x) =\int_{x}^{\infty} \frac{\left < \sigma v \right >}{x^2} dx .
\label{eq.RD}
\end{equation}
Here, $m_{\text{Pl}}= 1.22 \times 10^{19}$ GeV is the Planck mass, $g_{*}$
denotes the total number of relativistic degrees of freedom at
freeze-out\footnote{In the SM, $g_{*} \sim 120$ at $T \sim 1$ TeV and
  $g_{*} \sim 65$ at $T \sim 1$ GeV.}. $x_{f}=m_{T^0}/T_{f}$ is the
usual ratio of the DM mass to the freeze-out temperature $T_f$ and
$J(x)$ represents the efficiency of the post freeze-out
annihilation. In the non-relativistic limit, the thermal averaged
cross section has a linear dependence with the temperature as $ \left
< \sigma v \right > = a + 6 b \ T/m_{T^0}$ (here $T$ is the
temperature) , where $a$ and $b$ are terms describing the
non-relativistic cross section Taylor expansion in $v^{2}$ as $\sigma
\sim a + b v^{2}$ \cite{Griest:1990kh}.

For the total annihilation cross section there will be two contributions
in our case. First, there is the contribution from gauge interactions.
This case has been calculated in \cite{Cirelli:2005uq}, and for
the case of a hypercharge zero electro-weak triplet, the result
for the dominant annihilation channel to gauge bosons is simply:
\begin{equation}\label{sigvPG}
  \left < \sigma v \right > = \frac{g_2^4}{4 \pi m_{T^0}^2}.
\end{equation}
This cross section has the typical $m_{T^0}^{-2}$ dependence for a
thermal WIMP. Second, all nine operators, listed in table \ref{tab:ops6},
contribute to the DM annihilation. The cross section associated to any 
$d=6$ NRO can be written schematically as:
\begin{equation}
\sigma _{i} \sim \frac{C_{i}^2}{4 \pi \Lambda^{4}} f_{i} (m_{T^0},s,...),
\label{eq.sigv}
\end{equation}
where $f_{i}$ represents a function, specific to each operator, that
depends on kinematic variables, such as the DM mass and the center
of mass energy, $\sqrt{s}$, and so on.

We have used \texttt{CalcHEP} \cite{Belyaev:2012qa}, to calculate
$f_2$ and $f_3$ (corresponding to ${\cal O}_2$ and ${\cal O}_3$ from
table \ref{tab:ops6}). The result is:
\begin{equation}\label{fiOp}
  f_2 = a_2 m_q^2 \sqrt{\frac{(s - 4 m_q^2)}{(s - 4 m_{T^0}^2)}},
\; \hskip10mm
 f_3 = a_3 s  \sqrt{\frac{(s - 4 m_h^2)}{(s - 4 m_{T^0}^2)}},
\end{equation}
where $a_{2,3}$ are dimensionless numerical factors. Since in the
non-relativistic limit $s$ is given approximately as $s \sim 4
m_{T^0}^{2}(1+ v_{cm}^{2})$, with $v_{cm}$ the velocity of the DM
particle, one can approximate the $\sigma _{i}$ in the limit where
$m_q,m_h \ll m_{T^0}$ roughly as being $\sigma _{2} \propto
m_q^2/\Lambda^4$ and $\sigma _{3} \propto m_{T^0}^2/\Lambda^4$.

\begin{figure}[ht]
\centering
\includegraphics[width=0.48\textwidth]{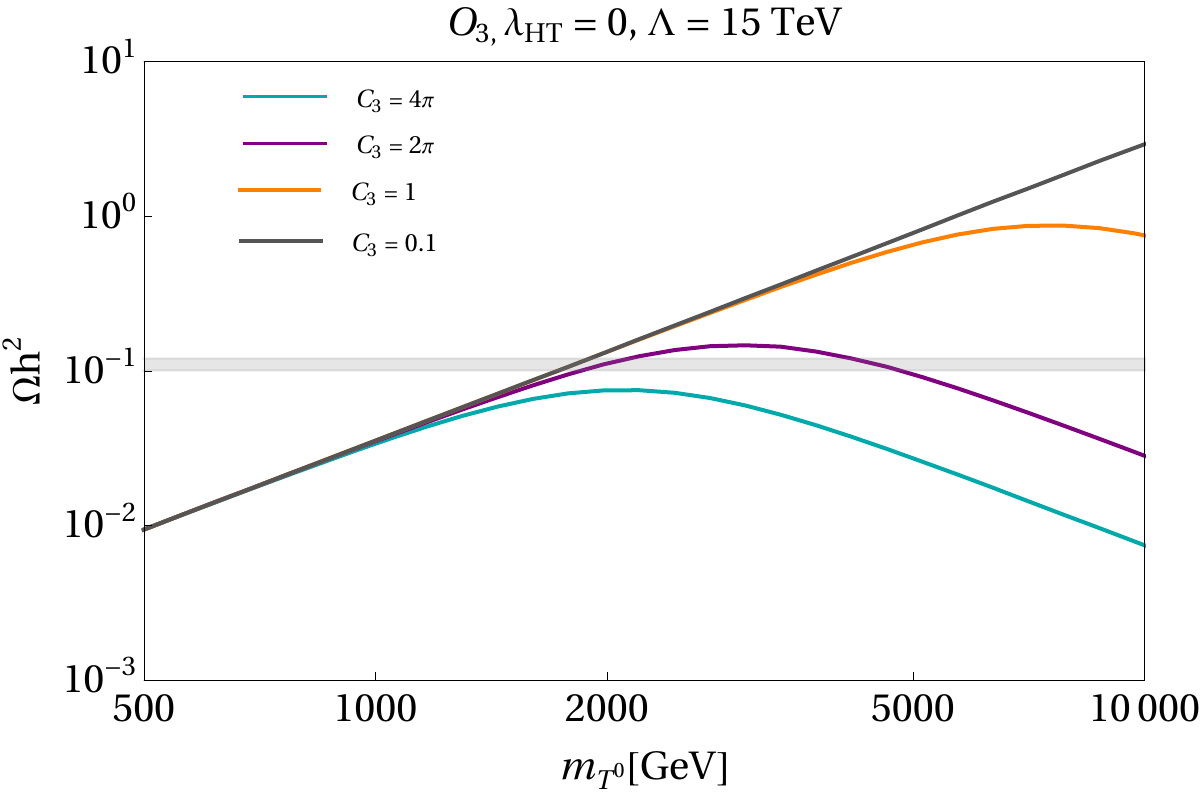}
\includegraphics[width=0.48\textwidth]{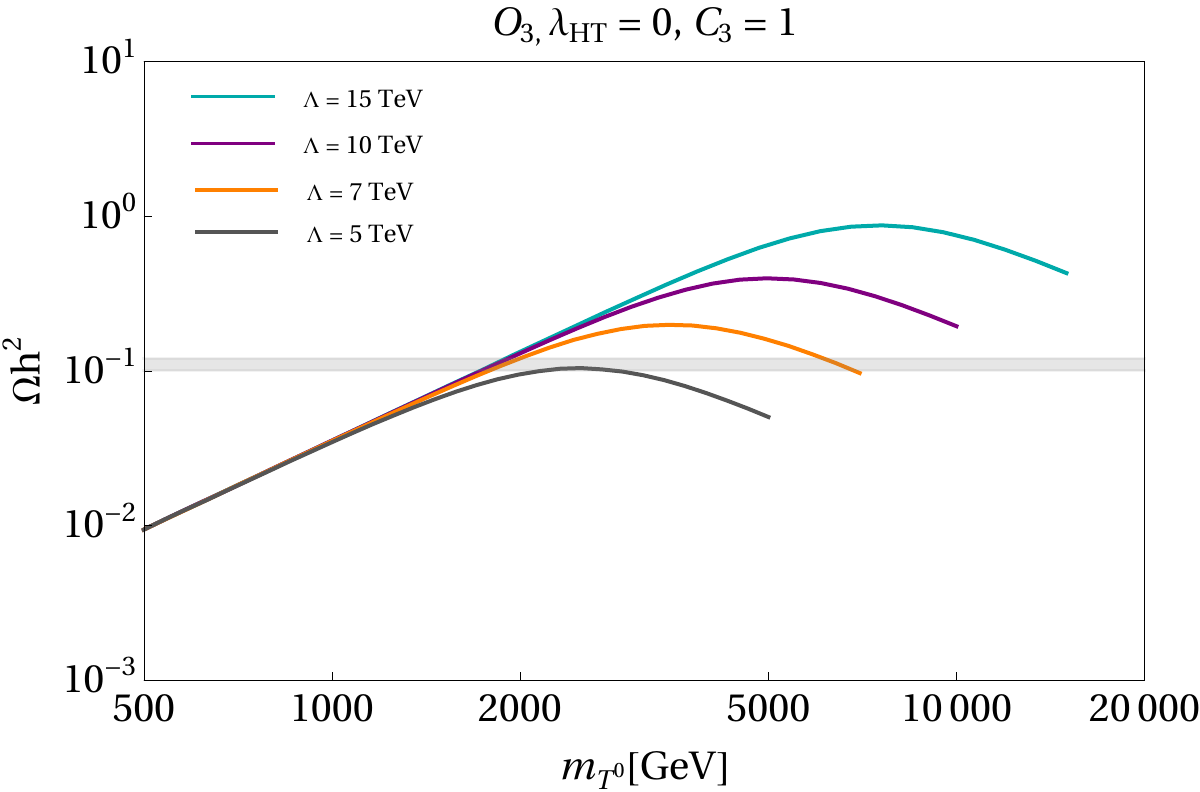}
\caption{Relic density as a funcion of $m_{T^0}$ (the mass of the dark
  matter candidate) with only $C_3$, the Wilson coefficient of ${\cal O}_{3}$,
  different from zero. To the left: For a fixed scale $\Lambda=15$ TeV
  and different values of $C_3$; to the right: For fixed $C_3=1$ and
  different values of the scale $\Lambda$. For this figure, we fixed
  $\lambda_{HT}=0$.}
\label{fig:rd}
\end{figure}

Let us first discuss the case of ${\cal O}_3$, see fig. \ref{fig:rd}.
The figure shows the calculated relic density, $\Omega h^2$, as
function of the mass of the dark matter candidate, for the case
$\lambda_{HT}=0$\footnote{We focus on the case $\lambda_{HT}=0$ in order to highlight the contribution of the effective operators. For the effect of $\lambda_{HT}\neq 0$ in the pure gauge scenario, see ref.~\cite{Katayose:2021mew,Bandyopadhyay:2024plc}.}, for different choices of $C_3$ (left) and for fixed
$C_3=1$ and different choices of $\Lambda$ (right). The grey band
indicates the allowed range of the DM density, as determined by PLANCK
\cite{Planck:2018vyg}, at $3\,\sigma\text{ C.L.}$

One can understand the observed behaviour with the help of the
equations given above. For smaller DM masses the calculated relic
density rises with increasing masses. This rise continues for all
values of $m_{T^0}$ for the smallest Wilson coefficient shown,
$C_3=0.1$ (on the left). Since $\Omega h^2 \propto 1/\sigma$,
eq. (\ref{eq.RD}), this behaviour directly reflects that for gauge
interactions $\sigma \propto 1/m_{T^0}^2$, see eq. (\ref{sigvPG}). For
larger values of $C_3$ the relic density starts to decrease again for
large values of $m_{T^0}$. This indicates the parameter region where
$\sigma \propto m_{T^0}^2/\Lambda^4$, i.e. where the contribution from
${\cal O}_3$ becomes dominant.

The plot on the right of fig. \ref{fig:rd} shows $\Omega h^2$ as function
of $m_{T^0}$ for different choices of $\Lambda$, for $C_3=1$. The
lines in this plot stop at $m_{T^0} \simeq \Lambda$, since we
use EFT to calculate the annihilation cross section. In fact, a consistent EFT
requires $m_{T^0} \ll \Lambda$, thus the calculation becomes less
reliable, when $m_{T^0}$ approaches $\Lambda$.\footnote{For UV models
which generate the operator via $t$-channel diagrams, departure from
EFT is expected to be of order $(m_{T^0}/\Lambda)^2$.} For values
of $\Lambda$ larger than $\Lambda  \sim 15$ TeV the contributions
from the operator will become sub-dominant relative to the gauge
interactions, unless we take $C_3$ larger than 1.

\begin{figure}[ht]
\centering
\includegraphics[width=0.48\textwidth]{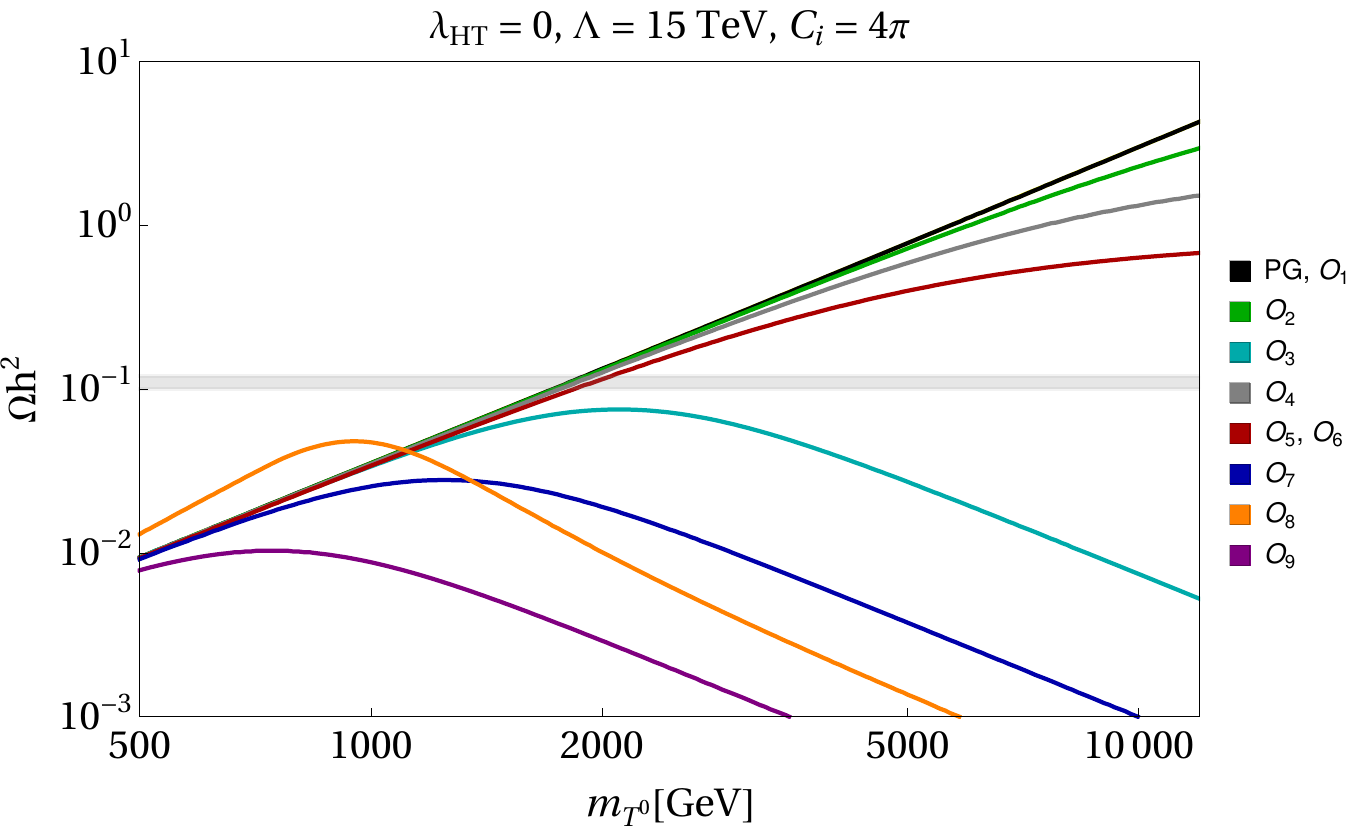}
\includegraphics[width=0.48\textwidth]{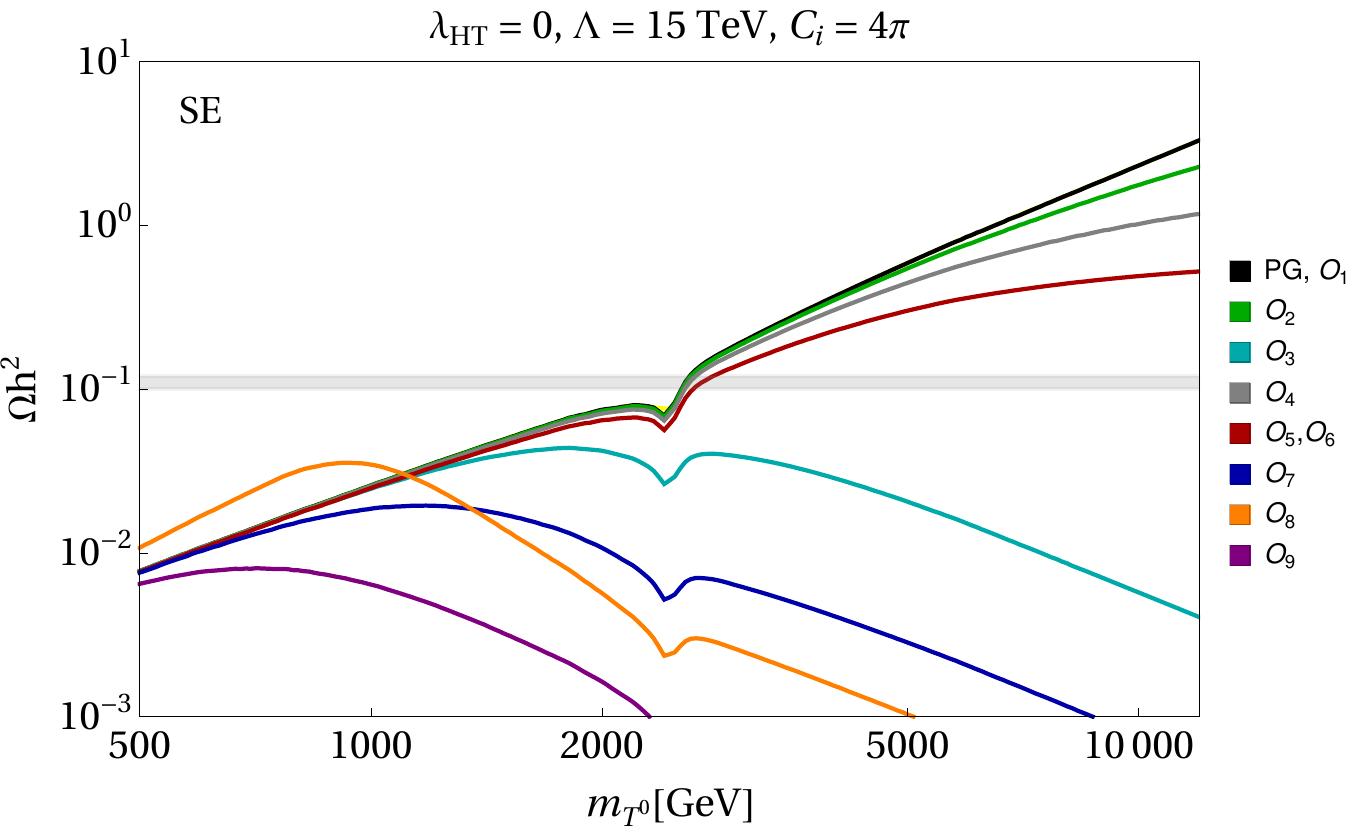}
\caption{Top panel: Relic density as a function of $m_{T^0}$ for the
  operators ${\cal O}_{1}- {\cal O}_{9}$ in table \ref{tab:ops6},
  ignoring (left) and including (right) the Sommerfield effect.}
\label{fig:rd2}
\end{figure}

Fig. \ref{fig:rd2} shows the relic density versus $m_{T^0}$ for the
first nine operators listed in table \ref{tab:ops6}. The plots contain
a black line labeled ``PG'', for the case of pure gauge interactions
(all operators switched off). In this figure, we show to the left
(right) the calculation without (including) the Sommerfeld effect.
In the calculation of the Sommerfeld effect, we follow the description
outlined in ref. \cite{Cirelli:2007xd}. These plots fix $\Lambda=15$ TeV
and all $C_i$ at their maximum value allowed by perturbativity
$C_i = 4\pi$.

One can clearly distinguish two groups of curves. There are the
operators ${\cal O}_{1,2,4,5,6}$ involving fermions and the operators
involving bosons, ${\cal O}_{3,7,8,9}$. The different behaviour can
be understood from eq. (\ref{fiOp}). Operators involving fermions
in the final state are chirally suppresed, i.e. $\sigma \propto
m_f^2$. These operators are therefore usually only a sub-dominant
contribution in the total annihilation cross section, compared
to the gauge interactions. Note that, operators involving fermions
carry generation indices, i.e. $C_k \to (C_k)^{ij}$, with
$i,j=1,2,3$. For this figure we switched on the coefficients
for all family indices, but due
to $\sigma \propto m_{f_{i}}^2$ the 3$^{\rm rd}$ generation indices
completely dominate the calculation.

Bosonic operators, on the other hand, lead to cross sections
$\sigma \propto m_{T^0}^2/\Lambda^4$. Thus, for ${\cal O}_{3,7,8,9}$
the relic density calculation will be completely dominated by
the operator contribution at larger values of $m_{T^0}$. In
section \ref{sec:NS} we will therefore concentrate the discussion
on the bosonic operators.

\section{Constraints from Direct and Indirect Detection}\label{sec:cnst}

In this section, we discuss the constraints from both direct and
indirect detection experiments on the effective operators $\mathcal
O_{1-9}$.

In the case of the direct detection, we derive the expression for the
cross section analytically.  The expression for the case of a scalar
$SU(2)$ triplet (without NROs) can be found in
Ref.~\cite{Katayose:2021mew}.  Here we add the contribution of the
different higher dimensional operators.

The spin-independent direct detection cross section for a DM particle
$T^0$ scattering off a nucleus $N$ is given by \cite{Chao:2018xwz}
\begin{align}
    \sigma_{SI} = \frac{\left|{\cal M}_{fi}\right|^2}{16 \pi(m_N+m_{T^0})^2},
\end{align}
where $m_N$, $m_{T^0}$ are the nucleus and dark matter particle
masses, respectively, and ${\cal M}_{fi}$ is the corresponding
amplitude:
\begin{align}
    {\cal M}_{fi} = \left<N T^0| {\cal L}_{\rm{int}} |N T^0\right>.
\end{align}
The operators from our effective Lagrangian (see table~\ref{tab:ops6})
that contribute to this amplitude at the tree level are those
involving quark currents ($\mathcal O_5$ and $\mathcal O_6$) and
gluons ($\mathcal O_9$).  Although $\mathcal O_2$ also involves
quarks, its contribution is negligible as it is proportional to the
square of the dark matter velocity ($v_{DM}^2$). The remaining
operators, which do not involve quarks or gluons, are all negligible
at the tree level.  Using the following expressions for the matrix
elements\footnote{These expressions use the non-relativistic
  normalization for nucleon one-particle states, while
  eqs.~\eqref{eq:NMEO5}-\eqref{eq:NMEO9} use relativistic normalization,
  i.e., we have introduced the corresponding replacements
  $\left|N\right>\rightarrow \sqrt{2E_N}\left|N\right> \approx
  \sqrt{2m_N}\left|N\right>$ when writing
  eqs.~(\ref{eq:NMEO5})-(\ref{eq:NMEO9}).}
\cite{Hisano:2010ct,Belanger:2013oya,Hisano:2015rsa}:

\begin{align}
\left<N\left|\frac{\alpha_S}{\pi}G_{\mu\nu}^a G^{a\mu\nu}\right|N\right> & =
-\frac{8}{9} f_{TG}^N m_N,\ \ \ f_{TG}^N=1-\sum_{q=u,d,s} f_{Tq}^N,\\
\left<N|m_q\,\bar q q |N\right> & = f_{Tq}^N m_N,
\end{align}
where $f_{Tu}^N=0.0153(0.011)$, $f_{Td}^N=0.0191(0.0273)$, $f_{Ts}^N=0.0447(0.0447)$ for the proton (neutron) \cite{Belanger:2013oya} and consequently $f_{TG}^N = 0.921(0.917)$, we find:
\begin{align}
    \left<N T^0\left| \frac{C_5^{ij} {\cal O}_5^{ij}}{\Lambda^2} \right|N T^0\right> & = 
    \frac{C_5^{ij}}{\Lambda^2} \left<N T^0\left| \overline{Q_i}d_{R_j} HTT \right|N T^0\right> = \frac{C_5^{ij}}{\Lambda^2}\frac{2v}{\sqrt{2}m_{d_i}} f_{d_i}^N \delta_{ij}\,2m_N^2,\label{eq:NMEO5}\\
    \left<N T^0\left| \frac{C_6^{ij} {\cal O}_6^{ij}}{\Lambda^2} \right|N T^0\right> & = 
    \frac{C_6^{ij}}{\Lambda^2} \left<N T^0\left| \overline{Q}u_RH^{\dagger}TT \right|N T^0\right> = \frac{C_6^{ij}}{\Lambda^2}\frac{2v}{\sqrt{2}m_{u_i}}  f_{u_i}^N \delta_{ij}\,2m_N^2,\label{eq:NMEO6}\\
    \left<N T^0\left| \frac{C_9 {\cal O}_9}{\Lambda^2} \right|N T^0\right> & = \frac{C_9}{\Lambda^2} \left<N T^0\left| G^{\mu\nu}G_{\mu\nu}TT \right|N T^0\right> = \frac{C_9}{\Lambda^2}\frac{4\pi}{\alpha_s}\left(-\frac{4}{9} f_{TG}^N\,2m_N^2\right),\label{eq:NMEO9}
\end{align}

Finally, the expression for the direct detection cross section including the effect of the effective operators is
\begin{align}
    \sigma_{\chi N} = \frac{\mu_{\chi N}^{2} m_{N}^2}{4\pi m_{\chi}^{2}} \left[\frac{2\lambda_{\chi H}}{m_{h}^{2}} f^N
    +\frac{C_5^{11}}{\Lambda^2} \frac{2v}{\sqrt{2}m_d} f^N_{Td}
    +\frac{C_5^{22}}{\Lambda^2} \frac{2v}{\sqrt{2}m_s} f^N_{Ts}
    +\frac{C_6^{11}}{\Lambda^2} \frac{2v}{\sqrt{2}m_u} f_{Tu}^N + \right.\nonumber
    \\
    \left.-\left(f_2^g+\frac{4\pi}{\alpha_s}\frac{C_9}{\Lambda^2}\right) \frac{4}{9}f_{TG}^N+\frac{3\pi\alpha_2^2}{4}\frac{m_{\chi}}{m_W^3}f_{\text{PDF}}^N \right]^2,\label{eq:ddxsec}
\end{align}
where $f^N = 0.287(0.284)$ and $f_{\text{PDF}}^N = 0.526(0.526)$ for proton (neutron)\cite{Katayose:2021mew}.

On the other hand, for the case of indirect detection bounds, we compute the gamma-ray flux arising from the annihilation of dark matter particles into gamma-rays, by using \texttt{MicrOMEGAs}. We will subsequently  study the complementarity of these predictions with direct detection searches, emphasizing the potential for detecting  
gamma-ray signals from each operator within the effective Lagrangian.

In this way, we will obtain the excluded regions of the parameter space for each operator by comparing the gamma-ray flux with the observational data from the H.E.S.S 
collaboration, and also the corresponding CTA excluded prospects. In all of them an Einasto profile~\cite{einasto1965construction} for the DM halo is assumed. 
Large uncertainties are expected  when other profiles are assumed, like the  Navarro, Frenk and White (NFW) 
profile~\cite{Navarro:1995iw,Navarro:1996gj}. Hence, the excluded and projected sensitivity regions in our results, could increase or decrease 
if other profiles for the dark matter halo are used.
For the excluded regions in the $W^+W^-$, $ZZ$ and $b\bar{b}$ channels 
we will use the data for 546~h from H.E.S.S~\cite{HESS:2022ygk},  and the CTA prospects from~\cite{CTA:2020qlo}. While for $\gamma\gamma$, we use 
the data for 546~h from H.E.S.S~\cite{Montanari:2023bzn} and the CTA prospects from \cite{CTAO:2024wvb}. 
Fermi-LAT bounds~\cite{atwood2009large} are slightly weaker than H.E.S.S for the explored regions of the parameter space, 
and are not used at all in our analysis.

\section{Numerical results\label{sec:NS}}

In this section we present our numerical results. We will study the
impact of the effective operators, see table \ref{tab:ops6},
considering one operator at a time. As discussed above, see sections
\ref{sec:tsmeft} and \ref{relic}, we can divide the operators in table
\ref{tab:ops6} in three groups. First there are the operators that
will be irrelevant for dark matter phenomenology, i.e.  ${\mathcal
  O}_{10-13}$.  The remaining operators we divide into (purely)
bosonic operators, $\mathcal O_{3,7,8,9}$, and operators with
fermions, $\mathcal O_{1,2,4,5,6}$. As shown in section \ref{relic},
the latter affect the relic density only mildly, thus from the
fermionic operators we will show only the results for one example
operator, i.e. ${\mathcal O}_6$.

In all cases discused in this section, the relic abundance of dark
matter and the indirect detection cross section calculation for each
operator was performed using \texttt{MicrOMEGAs}
\cite{Belanger:2013oya,Belanger:2018ccd,Belanger:2020gnr}, while the
direct detection cross section was obtained using the analytically
calculated cross sections given in section \ref{sec:cnst}, see eq. 
(\ref{eq:ddxsec}). We use the limits on direct detection cross
sections from LUX-ZEPLIN (LZ) experiment from
ref.~\cite{LZCollaboration:2024lux}.  Limits from XENON-nt are
similar, but currently slightly weaker \cite{XENON:2023cxc}. The
projected future bounds for DD were taken from the recent XLZD white
paper \cite{XLZD:2024nsu}. Note that the proposed DARWIN experiment
was expected to have a sensitivity very similar to XLZD (in its
40 ton baseline  configuration), see the projected DARWIN limits in
~\cite{Schumann:2015cpa}. For the indirect detection bounds we use
H.E.S.S. data from \cite{HESS:2022ygk,Montanari:2023bzn}, while
the future projections for CTA were taken from refs.
\cite{CTA:2020qlo,CTAO:2024wvb}.

\begin{figure}[h!]
\centering
\includegraphics[width=0.47\textwidth]{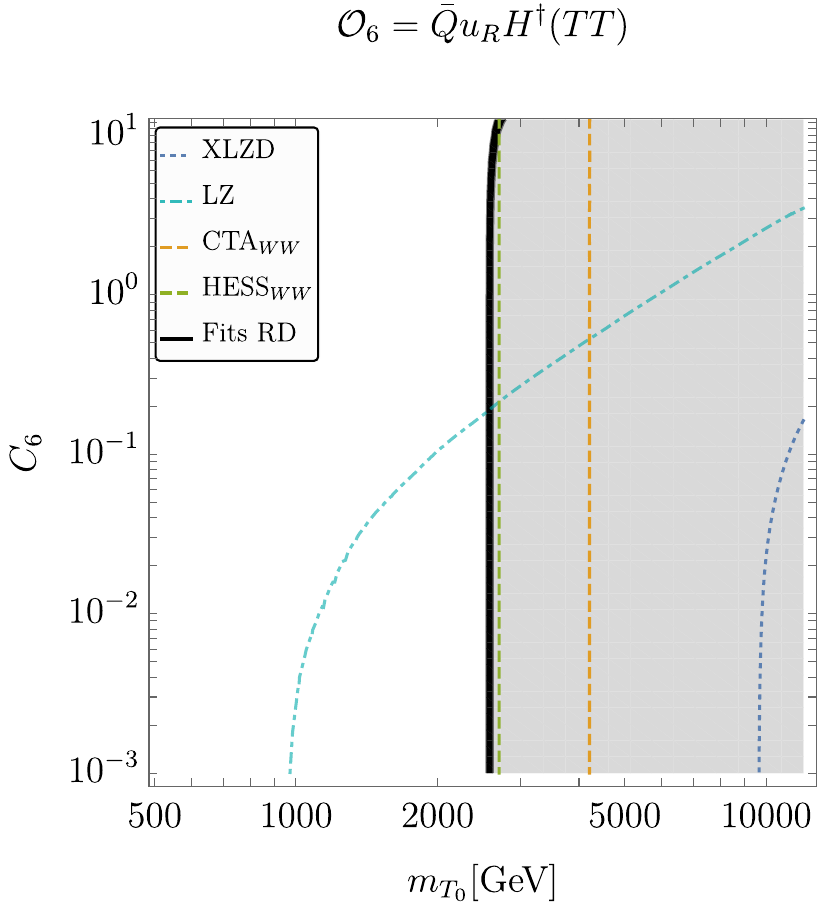}
\includegraphics[width=0.48\textwidth]{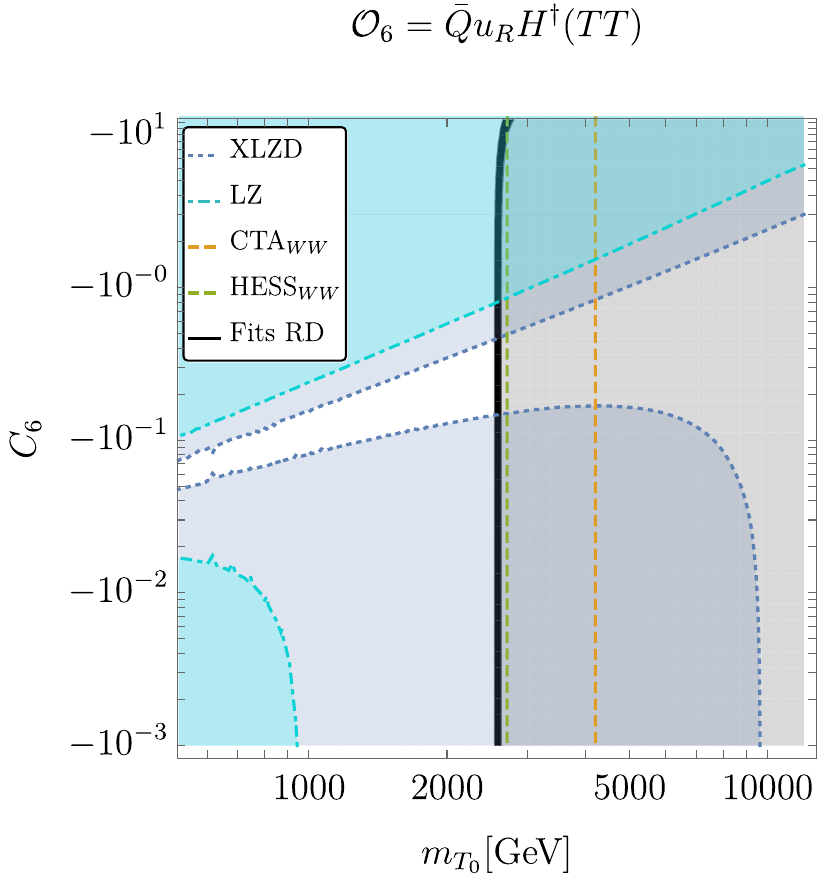}
\caption{Operator $\mathcal O_6$: Relic abundance, direct detection
and indirect detection bounds and prospects in the Wilson coefficient
  vs. DM mass plane.  \textit{\textbf{Left:}} For positive values of
  the Wilson coefficient.  \textit{\textbf{Right:}} For negative
  values of the Wilson coefficient.}
\label{fig:op6}
\end{figure}

\begin{figure}[h!]
\centering
\includegraphics[width=0.49\textwidth]{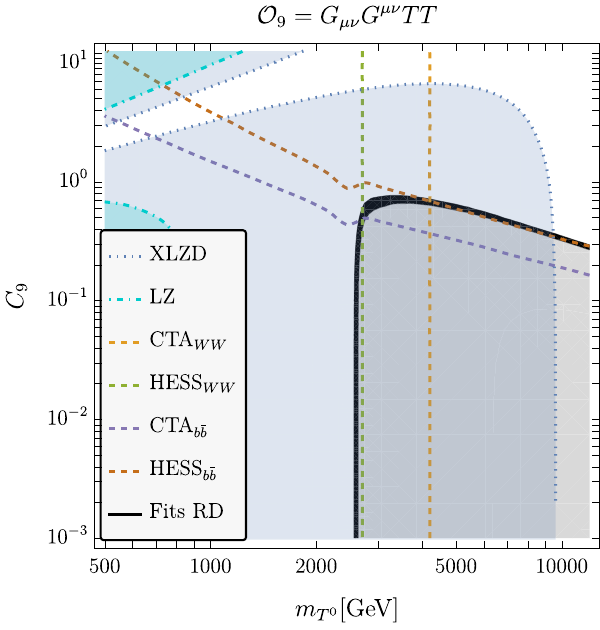}
\includegraphics[width=0.49\textwidth]{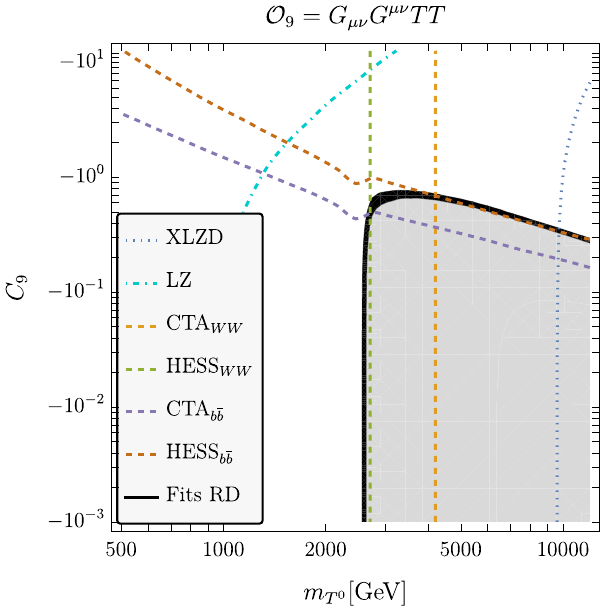}
\caption{Operator $\mathcal O_9$: Relic abundance, direct detection
  bounds, and indirect detection bounds in the Wilson coefficient
  vs. DM mass plane.  \textit{\textbf{Left:}} For positive values of
  the Wilson coefficient. Note that for both XLZD and LZ experiments
  here are two curves. The allowed region in each case corresponds to
  the area between these two curves.  \textit{\textbf{Right:}} For
  negative values of the Wilson coefficient.}\label{fig:op9}
\end{figure}

All figures in this section show the numerical results in the plane
Wilson coefficient, $C_i$, versus dark matter mass, $m_{T^0}$. As in fig.~\ref{fig:op6}, we
always set $\Lambda = 15\,\mbox{TeV}$ and $\lambda_{HT}=0$. The region
that reproduces correctly the relic density, as determined by PLANCK
\cite{Planck:2018vyg}, is shown by a solid black band. The grey area
to the right and below the black band indicates the excluded parameter
regions with an overabundant dark matter density.  Direct detection
constraints and prospects are shown as dash-dotted (LZ) or dotted (XLZD) lines respectively,
while indirect detection limits and prospects are marked with dashed lines. The most
dominant channels for indirect detection were considered for each
operator in the construction of the limits. These channels are
indicated in the label of each plot as a subscript under the
experiment name. In the figures, regions shown to the left or above
the limit curves are either excluded by current direct or indirect
detection experiments,\footnote{Note that some regions with masses
below $500\mbox{ GeV}$ could be allowed by some specific experiment
since both, direct and indirect detection bounds get weaker for small
masses.  However, this mass range is not relevant for us since it does
not reproduce the observed relic abundance and is therefore not shown
in the plots.}  or are projected to be within experimental sensitivity
for the case of future experiments. Additionally, for figures \ref{fig:op6} (right) and \ref{fig:op9} (left), for clarity we have shaded the excluded (or within future experimental sensitivity) areas, as the exclusion regions are more complex.

Let us discuss now the individual results in turn. We start with
${\mathcal O}_6$, our example fermionic operator. Fig.~\ref{fig:op6}
shows two plots for ${\mathcal O}_6$, to the left the case of
$C_6>0$, to the right $C_6<0$. As expected, the mass region which
fits correctly the relic density is independent of the operator, except
for $C_6$ very close to the non-perturbative limit, where a slight
increase in mass is seen. Since this is the same for all fermionic
operators, here we show only results for ${\mathcal O}_6$. 
For effective operators involving quarks, the direct detection cross
section is dominated by operators involving light quarks, while DM
annihilation (indirect detection) is dominated by heavy-quark
operators.  In fig.~\ref{fig:op6}, we consider the sum over all
three generations of quarks for all flavor-conserving Wilson
coefficients.

As the figure shows, ID constraints from H.E.S.S. already
significantly constrain the model. Taken at face value, $W^+W^-$ data seem
to exclude the mass region with the correct relic density. However,
recall that the constraints from H.E.S.S. use the Einasto profile and
as the collaboration states in \cite{HESS:2022ygk} other profiles
could weaken the limits by up to two orders of magnitude. The future
CTA experiment will improve sensitivity and very likely, provide a
decisive test for ${\mathcal O}_6$.
 
Also DD constraints rule out part of the available parameter space
already. Here, however, the discussion depends on the sign of the
Wilson coefficient.  Consider first the case $C_6>0$. Here, existing
DD constraints from LZ already rule out large values of $C_6$ in the
mass region, which correctly fits the relic density. XLZD will test
all the mass region that fits the relic density. For $C_6<0$ the
situation is more complicated. Note that here we have coloured
the region excluded (or to be tested) by LZ (XLZD), for an easier
comprehension of the constraints: The white region in the middle
of the plot will be left unconstrained even after XLZD. The rather
strange shape of the DD constraint for $C_6<0$ can be traced back
to a cancellation between the operator and the gauge contributions
to the DD cross section.

\begin{figure}[h!]
\centering
\includegraphics[width=0.49\textwidth]{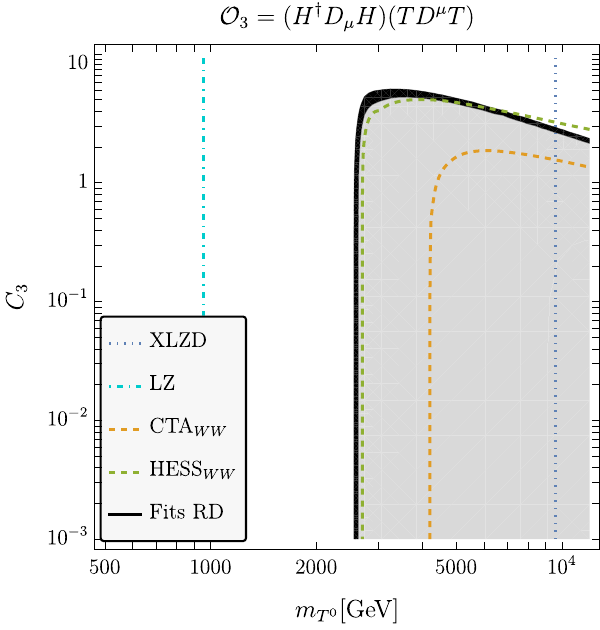}
\caption{Operator $\mathcal O_3$: Relic abundance, direct detection
  bounds, and indirect detection bounds in the Wilson coefficient
  vs. DM mass plane.  For the indirect detection bounds
  ($\text{CTA}_{WW}$ and $\text{HESS}_{WW}$) we have included both
  $W^+W^-$ and $ZZ$ annihilation modes since both are relevant for this
  operator and have similar gamma-ray spectra.}
\label{fig:op3}
\end{figure}

Fig.~\ref{fig:op3} shows the results for operator ${\mathcal O}_3$.
Here, we show only one plot, because results for positive and negative
Wilson coefficients are identical. For large values of $C_3$ there
is a new solution for explaining the correct relic density that
extends to quite large dark matter masses. In fact, we cut the plot 
at $m_{T^0}=15$ TeV, since $m_{T^0}\ge \Lambda$ is not allowed in
an EFT calculation, but solutions to fitting $\Omega h^2$ exist for
even larger masses (and larger values of $\Lambda$).

DD constraints on this operator are currently rather weak, but
XLZD will probe masses up to $m_{T^0}\approx 9.5$ TeV
in the future. H.E.S.S. results formally rule out the low-mass
region again, but allow $m_{T^0}\ge 7$ TeV even for the optimistic
Einasto profile. CTA will probe all mass regions, for which 
$\Omega h^2$ can be correctly fitted, at least for the more
favorable dark matter profiles.

\begin{figure}[h!]
\centering \includegraphics[width=0.48\textwidth]{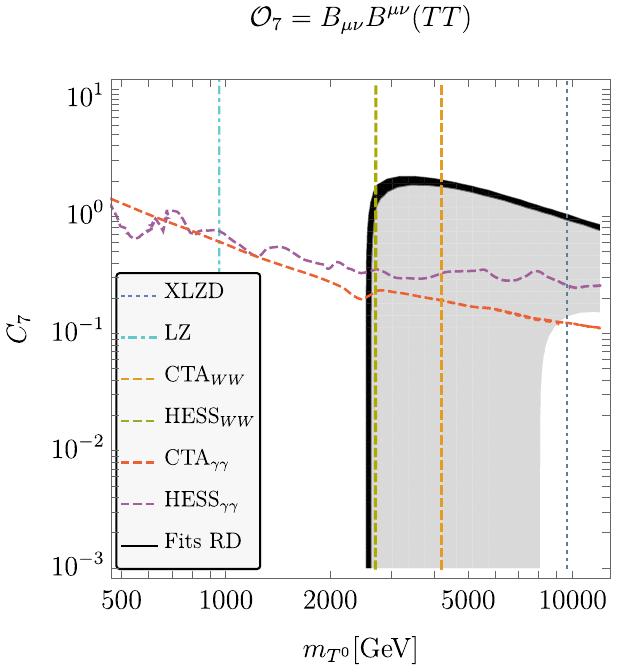}
\caption{Operator $\mathcal O_7$: Relic abundance, direct detection
  bounds, and indirect detection bounds in the Wilson coefficient
  vs. DM mass plane.}
\label{fig:op7}
\end{figure}

Results for $\mathcal O_7$ are shown in fig.~\ref{fig:op7}.  Results
are independent of the sign of $C_7$, as in the case of $\mathcal O_3$,
and also the parameter space for fitting the relic density is similar
to the one discussed above for $\mathcal O_3$. The main difference
between $\mathcal O_7$ and $\mathcal O_3$ is that $\mathcal O_7$ is
also constrained by searches of gamma-ray lines
($\text{HESS}_{\gamma\gamma}$ in the figure), which excludes the
region where the operator has an impact on the relic abundance
($C_7\gtrsim 7\times 10^{-1}$). Considering, additionally, the bounds
from $\text{HESS}_{WW}$, for this operator the whole region that fits
the relic abundance is excluded by indirect detection (for the
case of the Einasto profile). Thus, for $\mathcal O_7$ a sizeable
contribution of the operator to DM phenomenology is already now
disfavoured.

\begin{figure}[h!]
\centering
\includegraphics[width=0.49\textwidth]{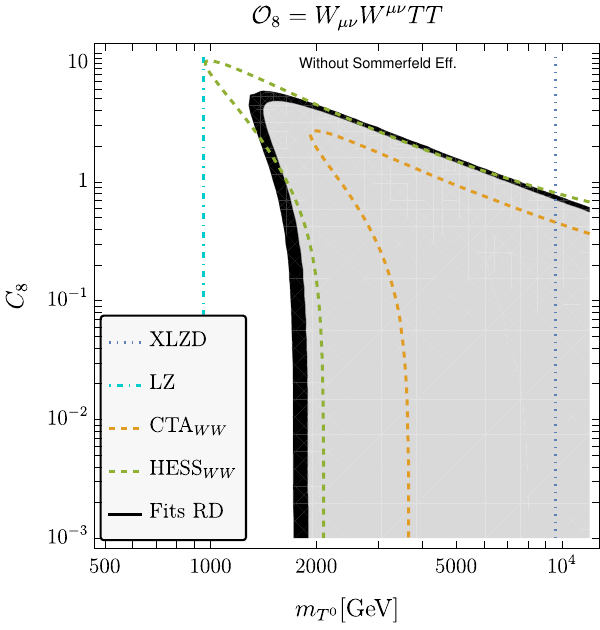}
\includegraphics[width=0.49\textwidth]{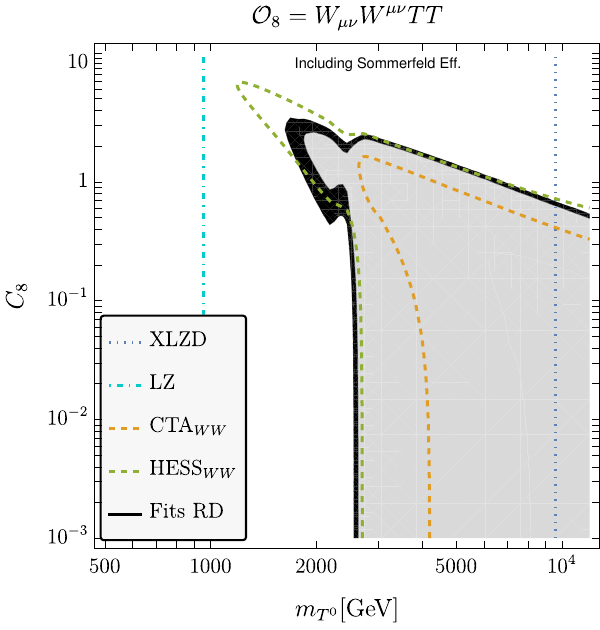}
\includegraphics[width=0.49\textwidth]{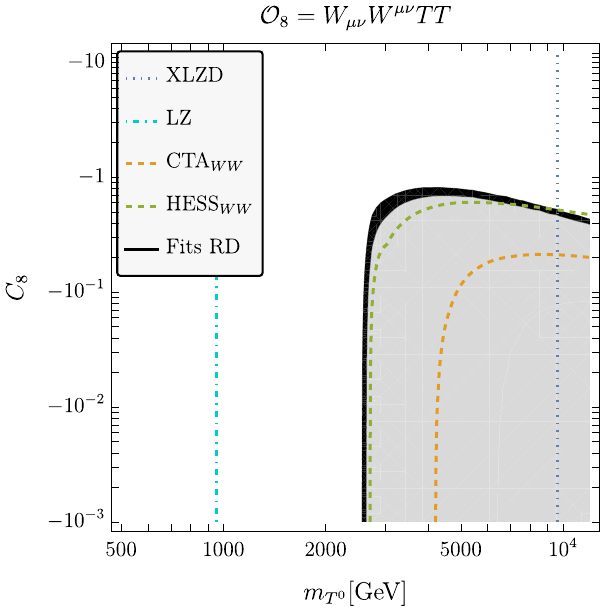}
\caption{Operator $\mathcal O_8$: Relic abundance, direct detection
  bounds, and indirect detection bounds in the Wilson coefficient
  vs. DM mass plane.  For the indirect detection bounds
  ($\text{CTA}_{WW}$ and $\text{HESS}_{WW}$) we have included both
  $W^+W^-$ and $ZZ$ annihilation modes since both are relevant in
  different mass ranges and have similar gamma-ray spectra ($ZZ$
  becomes important when the contribution of the operator interferes
  destructively with the pure gauge interactions.)
  \textit{\textbf{Upper left:}} Without Sommerfeld effect.
  \textit{\textbf{Upper right:}} Including Sommerfeld effect.
  \textit{\textbf{Bottom:}} For negative values of the Wilson
  coefficient.  }
\label{fig:op8}
\end{figure}

All the figures so far discussed include the effect of Sommerfeld
enhancement on the relic abundance and indirect detection bounds.  In
fig.~\ref{fig:op8}, upper panel, we show for illustration purposes both
results, with and without Sommerfeld effect, for the case $C_8>0$.
Note that, the calculation of the Sommerfeld enhancement for our
specific case of scalar triplet DM with zero hypercharge, was done in
ref. \cite{Cirelli:2007xd}, from where we extracted the enhancement
factor. The plot in the bottom panel of fig.~\ref{fig:op8} shows the
case $C_8<0$. Note that constraints from H.E.S.S. again allow only
particular mass and $C_8$ ranges (for the Einasto profile).  We can
see that for positive Wilson coefficients, there is a region that fits
the relic density and is consistent with current constraints for
relatively low DM masses ($1.7\,\mbox{TeV}\lesssim m_{T^0}\lesssim
2.5\,\mbox{TeV}$ and $2\lesssim C_8 \lesssim 4$). This region is
allowed due to interference of this operator with pure gauge
interactions and is not present for negative values of the Wilson
coefficient (as can be seen in the bottom panel in
fig.~\ref{fig:op8}).  Note also that large values of $m_{T^0} \gsim 8$
TeV remain allowed for both signs of $C_8$.

Finally, for $\mathcal O_9$, fig.~\ref{fig:op9}, we can see that the
region $2.6\,\mbox{TeV}\lesssim m_{T^0}\lesssim 3.5\,\mbox{TeV}$ (with
$C_9\approx \pm 7\times 10^{-1}$) is consistent with the observed relic
density and current constraints. This region will be probed in the
future by CTA and XLZD experiments.  Annihilation into pairs of $u$ or
$d$ quarks yields a gamma-ray spectrum similar to that of annihilation
into gluon pairs, while $c\bar{c}$ resembles $s\bar{s}$, and $ZZ$ is
analogous to $W^-W^+$ \cite{Fermi-LAT:2016afa}.  For this reason, and
just to give an example, in fig.~\ref{fig:op9}, we show also
indirect detection limits from CTA$_{b\bar{b}}$ and HESS$_{b\bar b}$,
because the gamma-ray spectrum of the $b\bar{b}$ channel is similar to
$gg$ \cite{Fermi-LAT:2016afa}, which is one of the dominant channels
in this case.

In summary, in this section we have discussed the impact of a set of
NRO $d=6$ operators on the relic density of $T^0$, as well as the
direct and indirect detection constraints on these operators. Future
indirect detection searches by CTA will provide an important test
for these operators.

\section{Conclusions}\label{sec:conclusions}

In this work we explored an extension of the standard model with a
scalar triplet dark matter candidate within an effective field theory,
which we denote as TSMEFT, i.e. the ``standard model effective field
theory supplemented by a scalar triplet''. We analyzed the effect of
non-renormalizable dimension-6 operators on the DM relic abundance,
the constraints on the parameter space from both direct and indirect
detection bounds and the prospects of testing the model with future
experiments.

The inclusion of $d=6$ operators can significantly affect the DM
phenomenology compared to the minimal model where only the
renormalizable interactions for scalar triplet are included
\cite{Cirelli:2005uq,Cirelli:2007xd}. $d=6$ operators can be classified
into two groups: (i) purely bosonic operators and (ii) operators with
fermions. For the bosonic operators we find that the relic density of
the triplet can be changed drastically. Since for these operators the
DM annihilation cross section is proportional to $\sigma \sim
m_{T^{0}}^{2}/\Lambda^{4}$, the operators can completely dominate the
relic density calculation at large values of $m_{T^0}$. This allows to
reproduce the experimental value of $\Omega h^2_{DM}$ for values of
$m_{T^0}$ in excess of 10 TeV, compared to $m_{T^0} \sim 2$ TeV ($2.8$
TeV) without (including) the Sommerfeld effect in the case of the
renormalizable model.  In contrast, operators involving fermions are
chirally suppresed, i.e $\sigma \sim m_{f}^2$, and thus contribute
only subdominantly to the the total annihilation cross section, when
compared with the gauge interactions. Fermionic operators therefore
are mostly irrelevant for fixing the relic density of the triplet.

Our main results are summarized in figures \ref{fig:op3}-\ref{fig:op9}.
These figures show the viable parameter space in the plane DM mass vs
Wilson coefficient, $C_i$, for the different operators. The plots show
the parameter combinations that reproduce the observed relic abundance
and add contours for current and future experimental constraints from
direct (LZ~\cite{LZCollaboration:2024lux}, XLZD~\cite{XLZD:2024nsu}) 
and indirect detection (H.E.S.S~\cite{HESS:2022ygk,Montanari:2023bzn},
CTA~\cite{CTA:2020qlo,CTAO:2024wvb}) experiments. All the figures include
the Sommerfeld effect for the relic abundance and the indirect
detection bounds.

Current DD bounds from LZ do not yet constrain the parameter space,
where the relic density is correctly fitted, with the exception of
${\cal O}_6$, for which LZ data provides a significant upper limit on
$C_6$ already. XLZD\footnote{ DARWIN is very
  similar to XLZD.}, on the other hand, will provide lower limits on
$m_{T^0}$ of the order of nearly ${\cal O}(10)$ TeV, thus either
ruling out large junks of the allowed parameter space or -- formulated
more optimistically - find dark matter, if TSMEFT is indeed the
correct model.

In contrast, indirect detection provides some important constraints on
the model already with current data. Searches by H.E.S.S in the $W^+W^-$,
$ZZ$ and $b\bar{b}$ channels ~\cite{HESS:2022ygk} are on the verge of
ruling out the pure renormalizable triplet model for masses that fit
$\Omega h^2$ correctly. Depending on the operator under
consideration, H.E.S.S allows still to fit $\Omega h^2$ correctly
for several operators, but mostly for larger triplet masses. The
improvements expected in the not-so-distant future from CTA data,
however, especially when combining $W^+W^-/ZZ$ and $\gamma\gamma$
channels, should lead to either a discovery of a DM signal or would
rule out TSMEFT as the correct explanation for dark matter.

Let us add a disclaimer. We need to stress again, that we have taken
both, the current ID data and the predictions for future experiments,
at face value.  Our optimistic conclusions depend strongly on the
assumption that the astrophysics (in particular the assumed dark
matter profiles) that went into the derivation of these limits are
correct. Nevertheless, we think TSMEFT is an attractive dark matter
model, since -- different from many other proposals -- it should be
possible to either discover DM signals or rule out the model
completely in the next decade or so.

\bigskip

\newpage
\centerline{\bf Acknowledgements}

\bigskip

We would like to thank Alexander Pukhov for useful discussions about \texttt{MicrOMEGAs}.
M.H. acknowledges support by Spanish grants PID2023-147306NB-I00 and
CEX2023-001292-S (MCIU/AEI/10.13039/501100011033), as well as
CIPROM/2021/054 (Generalitat Valenciana) and the MultiDark network,
RED2022-134411-T.
N.N. acknowledges support from ANID (Chile) FONDECYT Iniciaci\'on
Grant No. 11230879.
C.A. is supported by ANID-Chile FONDECYT grant No. 1231248 and
ANID-Chile PIA/APOYO AFB230003.
D.R. acknowledges support by Sostenibilidad UdeA, UdeA/CODI Grants 2022-52380 and 2023-59130, 
Minciencias Grants CD 82315 CT ICETEX 2021-1080. M.G acknowledges support from Centro
de F\'isica Te\'orica de Valpara\'iso (CeFiTeV) and project PFE UVA22991/PUENTE.
\bigskip

\appendix
\section{Appendix}\label{sec:appendix}

\subsection{$T$-odd operators}
\label{subsec:OddOps}

In this appendix we give for completenss also the operators 
that are odd in $T$ up to $d=6$. All operators
are given in table~\ref{tab:OddOps}. As in table~\ref{tab:ops6},
we specify the fields involved in the operator and give the
parameter counting, without spelling out the generation
indices explicitly. If any of these operators is present in
the theory, $T$ will be unstable.

\begin{table}
\centering
\begin{tabular}{cccc}
\hline
Name & Operator & $\#$ real parameters & +h.c.? \\
\hline
\hline
${\cal O}_{WBT}$ & $W^{\mu\nu} B_{\mu\nu}T$ & 1   & no  \\
${\cal O}_{LeHT}$ & $\overline{L}e_RHT$   & 18   & yes \\
${\cal O}_{QdHT}$ & $\overline{Q}d_RHT$   & 18   & yes \\
${\cal O}_{QuHT}$ & $\overline{Q}u_RH{^\dagger}T$   & 18   & yes \\
${\cal O}_{H^4T}$ & $H^{\dagger}H^{\dagger}HHT$  & 1   & no  \\
${\cal O}_{H^2T^3}$ & $H^{\dagger}HTTT$      & 1   & no  \\
\hline
${\cal O}_{L^2H^2T}$ & $LLHHT$             & 18   & yes  \\
\hline
\end{tabular}
\caption{Operators odd in $T$ in TSMEFT. All operators in this table are
  $d=5$, except the last one which is $d=6$. These operators need to
  be eliminated, if $T$ is to be a good DM candidate. }
\label{tab:OddOps}
\end{table}

\subsection{Operators for complex $T$}
\label{subsec:CmplOps}

In this appendix we discuss for completenss also operators up to $d=6$
that are non-zero only if $T$ is assumed to be a complex field.
Since $T$ has hypercharge zero, it can be a self-conjugate field. 
We work with this assumption throughout the paper. 

However, one could also assume that $T$ is complex and for that 
case more Lagrangian terms are allowed. For example, 
there are two types of mass terms for a complex $T$:
\begin{equation}\label{eq:LagC}
m_C^2 |T|^2  + \frac{1}{2} m_R^2 ( T^2 + {\rm h.c.})
\end{equation}
Similarly, for the non-renormalizable operators for complex $T$ 
there are many more independent contractions. For example, the 
single $T^6$ will become 4 different terms: $T^6$, $T^{\dagger}T^5$ 
$(T^{\dagger})^2T^4$ and $(T^{\dagger})^3T^3$. Most of the time, 
these new operators do not play any significant role for the 
dark matter phenomenology of $T$, but there are three possible 
exceptions:
\begin{equation}\label{eq:OpC}
{\cal O}_{e^2T^2}=(\overline{e_R}\gamma_{\mu}e_R)(T^{\dagger}D^{\mu}T),
\quad
{\cal O}_{d^2T^2}=(\overline{d_R}\gamma_{\mu}d_R)(T^{\dagger}D^{\mu}T),
\quad
{\cal O}_{u^2T^2}=(\overline{u_R}\gamma_{\mu}u_R)(T^{\dagger}D^{\mu}T).
\end{equation}
These operators, vanishing for real $T$, do have a non-zero effect on
the relic density similar to ${\cal O}_{L^2T^2}$ and 
${\cal O}_{Q^2T^2}$, that we discuss in the main text.

\subsection{Tree-level UV completions for TSMEFT operators\label{subsec:UV}}

In this appendix we give the tree-level decomposition for $d=6$ TSMEFT
operators with even number of $T$. From the 9 operators studied in
detail in this paper only the first six operators given in table~\ref{tab:ops6}
can be decomposed at tree-level. The remaining 
three operators, ${\cal O}_{7}-{\cal O}_{9}$, can be decomposed only
at 1-loop.

Table~\ref{tab:decomp1} gives the decompositions for
${\cal O}_{1}-{\cal O}_{3}$. These operators can be decomposed adding
only one additional BSM field. The table indicates whether the
internal field in the diagram is a vector ($V$) or fermion ($F$) and
the transformation properties under the SM gauge group. Also,
depending on the decomposition the internal particle needs to be
either odd or even under a $Z_2$, this is given as $-1$ or $1$ in the
last index of the field.

\begin{table}[h]
\centering
\begin{tabular}{cccc}
\hline
Name & Operator & Fields  \\
\hline
\hline
${\cal O}_{1}$ & $(\overline{L}\gamma_{\mu}L)(TD^{\mu}T)$ &
$V_{1,1,0,1}$, $V_{1,3,0,1}$, $F_{1,2,1/2,-1}$, $F_{1,4,1/2,-1}$  \\
${\cal O}_{2}$ & $(\overline{Q}\gamma_{\mu}Q)(TD^{\mu}T)$ &  
$V_{1,1,0,1}$, $V_{1,3,0,1}$, $F_{3,2,1/6,-1}$, $F_{3,4,1/6,-1}$  \\
${\cal O}_{3}$ & $(H^{\dagger}D_{\mu}H)(TD^{\mu}T)$  &
$S_{1,1,0,1}$, $S_{1,3,0,1}$, $S_{1,2,1/2,-1}$, $S_{1,4,1/2,-1}$, \\
 & & 
$V_{1,1,0,1}$, $V_{1,3,0,1}$, $V_{1,2,1/2,-1}$, $V_{1,4,1/2,-1}$  \\
\hline
\end{tabular}
\caption{Tree-level decompositions for operators even in $T$ in TSMEFT
  at $d=6$. Fields are given with their transformation properties
  under $SU(3)_c\times SU(2)_L \times U(1)_Y \times Z_2$. The last
  index indicates even or odd under $Z_2$. This table gives operators
  that can be decomposed with one BSM field.}
\label{tab:decomp1}
\end{table}

\begin{figure}[t]
    \centering
    \includegraphics[scale=0.6]{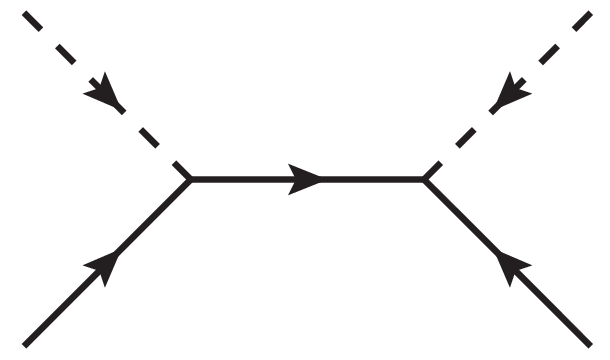} \hskip8mm
    \includegraphics[scale=0.6]{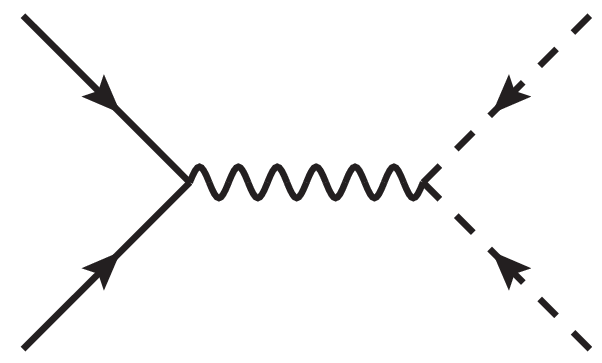} \\ \vskip4mm
    \includegraphics[scale=0.6]{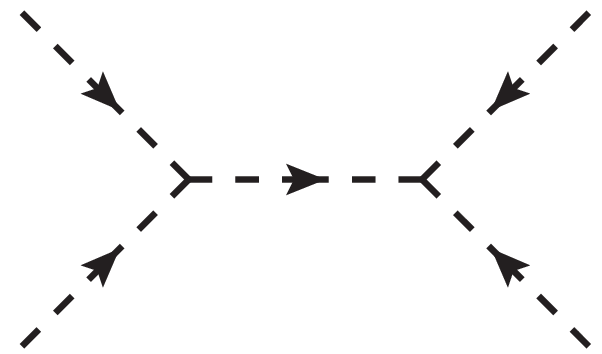} \hskip8mm
    \includegraphics[scale=0.6]{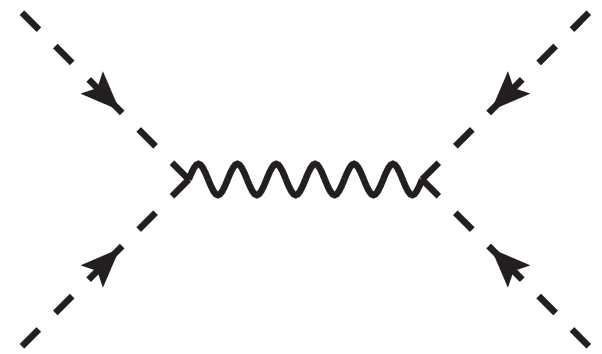} \\ \vskip4mm
    \includegraphics[scale=0.45]{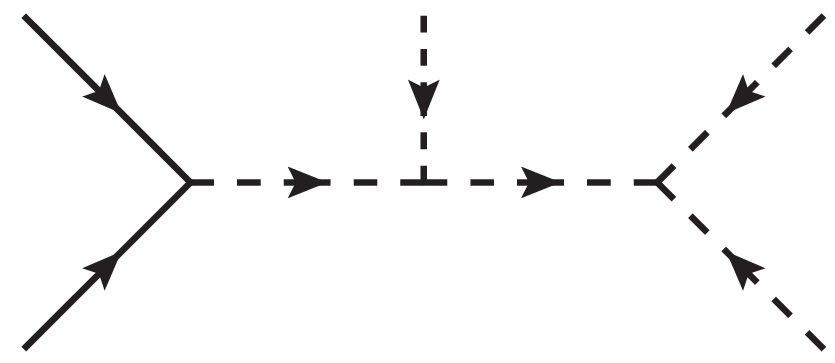} \hskip8mm
    \includegraphics[scale=0.45]{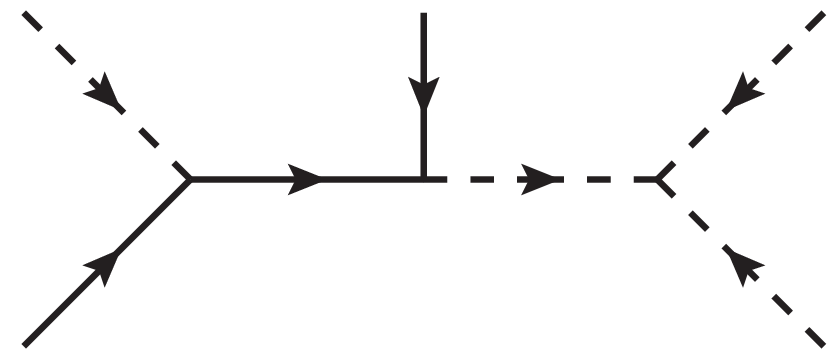} \\ \vskip2mm
    \includegraphics[scale=0.45]{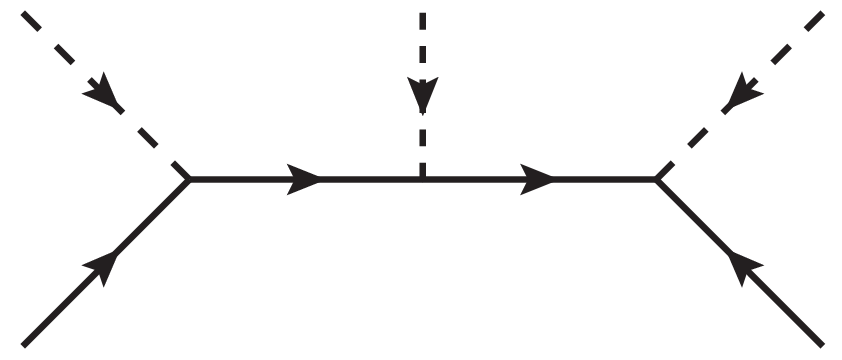} \hskip8mm
    \includegraphics[scale=0.45]{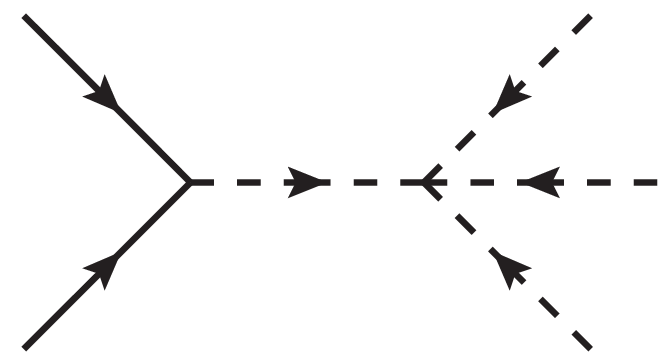}
    \caption{Diagrams for the decomposition of operators ${\cal O}_{1}-
      {\cal O}_{6}$. Top line: ${\cal O}_{1,2}$; 2nd line: ${\cal O}_{3}$.
      The bottom two lines are for ${\cal O}_{4}-{\cal O}_{6}$. Here,
      three diagrams need two different BSM fields.}
    \label{fig:UVDiags}
\end{figure}

Table~\ref{tab:decomp2} gives the decompositions for operators
${\cal O}_{4}-{\cal O}_{6}$. As can be seen from
fig.~\ref{fig:UVDiags}, there are four possible diagrams for these
``Yukawa-like'' operators and all except the last diagram need two BSM
fields to generate the operator.

\begin{table}[h]
\scalebox{0.9}{
\begin{tabular}{cccc}
\hline
Name & Operator & Fields  \\
\hline
\hline
${\cal O}_{4}$ & $\overline{L}e_RHTT$ &
($S_{1,2,1/2,1},S_{1,2,1/2,-1}$), ($S_{1,2,1/2,1},S_{1,4,1/2,-1}$),
($S_{1,2,1/2,1},S_{1,1,0,1}$), ($S_{1,2,1/2,1},S_{1,3,0,1}$), \\ & & 
($F_{1,2,1/2,1},F_{1,2,1/2,-1}$), ($F_{1,2,1/2,1},F_{1,4,1/2,-1}$),
($F_{1,3,1,-1},F_{1,1,1,1}$), ($F_{1,3,1,-1},F_{1,3,1,1}$), \\ & & 
($F_{1,3,1,-1},F_{1,2,1/2,-1}$), ($F_{1,3,1,-1},F_{1,4,1/2,-1}$),
($F_{1,2,1/2,1},S_{1,1,0,1}$), ($F_{1,2,1/2,1},S_{1,3,0,1}$),\\ & & 
($F_{1,3,1,-1},S_{1,2,1/2,-1}$), ($F_{1,3,1,-1},S_{1,4,1/2,-1}$),
($F_{1,1,1,1},S_{1,1,0,1}$), ($F_{1,3,1,1},S_{1,3,0,1}$), \\ & & 
($F_{1,2,1/2,-1},S_{1,2,1/2,-1}$),($F_{1,4,1/2,-1},S_{1,4,1/2,-1}$),
$S_{1,2,1/2,1}$ \\
${\cal O}_{5}$ & $\overline{Q}d_RHTT$ &
($S_{1,2,1/2,1},S_{1,2,1/2,-1}$), ($S_{1,2,1/2,1},S_{1,4,1/2,-1}$),
($S_{1,2,1/2,1},S_{1,1,0,1}$), ($S_{1,2,1/2,1},S_{1,3,0,1}$), \\ & & 
($F_{3,2,1/6,1},F_{3,2,1/6,-1}$), ($F_{3,2,1/6,1},F_{3,4,1/6,-1}$),
($F_{3,3,1/3,-1},F_{3,1,1/3,1}$), ($F_{3,3,1/3,-1},F_{3,3,1/3,1}$), \\ & & 
($F_{3,3,1/3,-1},F_{3,2,1/6,-1}$), ($F_{3,3,1/3,-1},F_{3,4,1/6,-1}$),
($F_{3,2,1/6,1},S_{1,1,0,1}$), ($F_{3,2,1/6,1},S_{1,3,0,1}$),\\ & & 
($F_{3,3,1/3,-1},S_{1,2,1/2,-1}$), ($F_{3,3,1/3,-1},S_{1,4,1/2,-1}$),
($F_{3,1,1/3,1},S_{1,1,0,1}$), ($F_{3,3,1/3,1},S_{1,3,0,1}$), \\ & & 
($F_{3,2,1/6,-1},S_{1,2,1/2,-1}$),($F_{3,4,1/6,-1},S_{1,4,1/2,-1}$),
$S_{1,2,1/2,1}$ \\
${\cal O}_{6}$ & $\overline{Q}u_RH^{\dagger}TT$ &
($S_{1,2,1/2,1},S_{1,2,1/2,-1}$), ($S_{1,2,1/2,1},S_{1,4,1/2,-1}$),
($S_{1,2,1/2,1},S_{1,1,0,1}$), ($S_{1,2,1/2,1},S_{1,3,0,1}$), \\ & & 
($F_{3,2,1/6,1},F_{3,2,1/6,-1}$), ($F_{3,2,1/6,1},F_{3,4,1/6,-1}$),
($F_{3,3,2/3,-1},F_{3,1,2/3,1}$), ($F_{3,3,2/3,-1},F_{3,3,2/3,1}$), \\ & & 
($F_{3,3,2/3,-1},F_{3,2,1/6,-1}$), ($F_{3,3,2/3,-1},F_{3,4,1/6,-1}$),
($F_{3,1,2/3,1},S_{1,1,0,1}$), ($F_{3,1,2/3,1},S_{1,3,0,1}$),\\ & & 
($F_{3,2,1/6,-1},S_{1,2,1/2,-1}$), ($F_{3,2,1/6,-1},S_{1,4,1/2,-1}$),
($F_{3,2,1/6,1},S_{1,1,0,1}$), ($F_{3,4,1/6,1},S_{1,3,0,1}$), \\ & & 
($F_{3,3,2/3,-1},S_{1,2,1/2,-1}$),($F_{3,3,2/3,-1},S_{1,4,1/2,-1}$),
$S_{1,2,1/2,1}$ \\
\hline
\end{tabular}
}
\caption{Tree-level decompositions for operators even in $T$ in TSMEFT
  at $d=6$: Yukawa-like operators. This table gives operators
  that need two BSM fields, except for the last decomposition
corresponding to diagram in the bottom right of fig.~\ref{fig:UVDiags}.}
\label{tab:decomp2}
\end{table}

\bibliographystyle{JHEP}
\bibliography{TSMEFT.bib}

\end{document}